\documentclass[aps,prb,preprint,superscriptaddress,floatfix,altaffilletter]{revtex4}

\usepackage{graphicx}
\usepackage{dcolumn}
\usepackage{bm}
\usepackage{amsmath}
\usepackage{multirow}
\usepackage{hhline}

\newcommand{\tbhc}[1]{\multicolumn{1}{c}{\textbf{#1}}}
\newcommand{\tbhl}[1]{\multicolumn{1}{l}{\textbf{#1}}}
\newcommand{\tbhr}[1]{\multicolumn{1}{r}{\textbf{#1}}}
\newcommand{\tbr}[1]{\multicolumn{1}{r}{#1}}
\newcommand{\tbc}[1]{\multicolumn{1}{c}{#1}}

\newcolumntype{f}[1]{D{.}{.}{#1}}

\renewcommand{\deg}[0]{$^{\circ}$}

\begin{document}

\title{Basis set effects on the hyperpolarizability of CHCl$_3$: Gaussian-type
orbitals, numerical basis sets and real-space grids}

\date{\today}
\author{Fernando D. Vila}
\affiliation{Department of Physics, University of Washington, Seattle, WA 98195}
\author{David A. Strubbe}
\affiliation{Department of Physics, University of California, Berkeley, and
Materials Sciences Division, Lawrence Berkeley National Laboratory}
\author{Yoshinari Takimoto}
\affiliation{Institute for Solid State Physics, University of
Tokyo, Kashiwa, Chiba 277-8581, Japan}
\affiliation{Department of Physics, University of Washington, Seattle, WA 98195}
\author{Xavier Andrade}
\affiliation{Nano-Bio Spectroscopy group and ETSF Scientific Development Centre,
Dpto. F\'\i sica de Materiales, Universidad del Pa\'\i s  Vasco, Centro de
F\'\i sica de Materiales CSIC-UPV/EHU-MPC and DIPC, Av. Tolosa 72,
E-20018 San Sebasti\'an, Spain}
\author{Angel Rubio}
\affiliation{Nano-Bio Spectroscopy group and ETSF Scientific Development Centre,
Dpto. F\'\i sica de Materiales, Universidad del Pa\'\i s  Vasco, Centro de
F\'\i sica de Materiales CSIC-UPV/EHU-MPC and DIPC, Av. Tolosa 72,
E-20018 San Sebasti\'an, Spain}
\affiliation{Fritz-Haber-Institut der Max-Planck-Gesellschaft, Berlin, Germany}
\author{Steven G. Louie}
\affiliation{Department of Physics, University of California, Berkeley, and
Materials Sciences Division, Lawrence Berkeley National Laboratory}
\author{John J. Rehr}
\affiliation{Department of Physics, University of Washington, Seattle, WA 98195}

\begin{abstract}

Calculations of the hyperpolarizability are typically much more difficult to
converge with basis set size than the linear polarizability.
In order to understand these convergence issues and hence obtain accurate
\textit{ab initio} values, we compare
calculations of the static hyperpolarizability of the gas-phase chloroform
molecule (CHCl$_3$) using three different kinds of basis sets: Gaussian-type
orbitals, numerical basis sets, and real-space grids. Although all of these
methods can yield similar results, surprisingly large, diffuse basis sets are
needed to achieve convergence to comparable values. These results are
interpreted in terms of local polarizability and hyperpolarizability densities.
We find that the hyperpolarizability is very sensitive to the molecular
structure, and we also assess the significance of vibrational contributions and
frequency dispersion. 

\end{abstract}

\date{\today}

\maketitle

\section{Introduction}
\label{sec:intro}

Chloroform (CHCl$_3$) is a widely used solvent in measurements of nonlinear
optical properties of organic chromophores, using techniques such as
electric-field-induced second-harmonic generation (EFISH) and hyper-Rayleigh
scattering (HRS).\cite{kaatz98, Das2006} It is sometimes also used as an
internal reference.\cite{clays} However, assumptions have to made to extract
molecular hyperpolarizabilities from these measurements, in particular from EFISH which
actually measures a third-order response function.
For calibration purposes, either absolute measurements
or \textit{ab initio} calculations are needed to convert between the different
combinations of tensor components of the hyperpolarizability measured in the
EFISH and HRS experiments. However, very early calculations, which have been
recognized as unsatisfactory by their authors,\cite{Karna1990} have heretofore
been used for such conversions.\cite{kaatz98} Consequently there is a need
for better understanding of the convergence issues and for
more accurate \textit{ab initio} calculations. Toward this end 
we have carried out a systematic study of the second-order hyperpolarizability
$\beta$ of chloroform using several theoretical methods. In
an effort to obtain high-quality results which improve on earlier
calculations, we have based our calculations on an accurate experimental
structure and also considered both frequency-dependence and vibrational
contributions, effects which are typically neglected in other calculations. 

Although our presentation
is restricted to a single system, the methodology is more general and
hence many of our results will likely be applicable to many other cases.
Likewise, the quantitative comparison of several theoretical methods
is both novel and of general interest.  Finally we believe our
interpretation of the linear and nonlinear response in terms of local
polarizability densities provides a useful way of understanding the local
contributions to the polarizability from various parts of a molecule.

Chloroform is of particular theoretical interest because its hyperpolarizability
is challenging to measure experimentally due its small magnitude
compared to typical experimental errors, and hence
the available measurements have both positive and negative values, with
large error bars.\cite{kaatz98, Miller1977} Similarly,
this nonlinear property has proved to be quite difficult to calculate
theoretically, as the results exhibit a large dependence on the quality of
the basis set used, both for DFT and coupled-cluster
methods,\cite{davidson06, Hammond} as well as the molecular geometry.
One of the main purposes of 
this paper is to investigate the reasons for these difficulties using three
different basis set approaches: i) Gaussian-type orbitals (GTOs); ii) numerical
basis sets; and iii) real-space grids, all with comparable treatments of 
exchange and correlation.  The importance of different aspects of the basis sets
(diffuseness, polarization, etc.) was studied systematically by changing the
number of GTOs, the cutoff radii of the numerical basis sets, and the extent
and density of the real-space grids. In order to interpret these results, we
also studied the spatial distribution of the dielectric properties using the
concepts of polarizability- and hyperpolarizability-densities, as well as first-
and second-order electric-field-perturbed densities.\cite{chopra89,nakano05} In
addition, we also investigated the dependence of polarization properties on the
choice of exchange-correlation (XC) functionals and the correlation level by
means of DFT, HF, MP2 and CCSD methods. We also briefly discuss the dependence
of the results on the molecular geometry, which was found to have a significant influence on the calculated hyperpolarizability.

Unless noted explicitly, the experimental molecular geometry of Colmont et
al.\cite{Colmont1998158} was used throughout this work: $r_{\mathrm{CH}}=1.080$
\AA, $r_{\mathrm{CCl}}=1.760$ \AA\ and $\angle{\mathrm{HCCl}}=108.23$\deg. The
molecule was located with its center of mass at the origin, and oriented with
the CH bond along the positive $z$-direction and one HCCl angle in the
$yz$-plane. Since chloroform has C$_{3v}$ point-group symmetry, the following
symmetry relations hold for the linear polarizability $\alpha$ and hyperpolarizability $\beta$: $\alpha_{xx} = \alpha_{yy}$, $\beta_{xxy} =
-\beta_{yyy}$ and $\beta_{xxz} = \beta_{yyz}$. In the static case Kleinman
symmetry\cite{kleinman62} also applies. Thus the $\alpha_{yy}$, $\alpha_{zz}$,
$\beta_{yyy}$, $\beta_{yyz}$ and $\beta_{zzz}$ components fully describe the
polarizability and hyperpolarizability tensors; all other permutations of the
indices are equivalent. In the dynamic case at non-zero frequency, however, the
components of $\beta_{ijk} \left( -2 \omega; \omega, \omega \right)$ are not all
equivalent: $\beta_{yyz} = \beta_{yzy} \neq \beta_{zyy}$. Here we use the
Taylor convention for hyperpolarizabilities.\cite{willetts92}

From our calculated tensor components we can calculate the
isotropically averaged polarizability $\bar{\alpha} = \frac{1}{3} \sum_{i}
\alpha_{ii}$, the second-order hyperpolarizability coefficient, and
the EFISH hyperpolarizability in the direction of the dipole
moment $\beta_{\parallel i} = \frac{1}{5} \sum_j \left(\beta_{ijj} + \beta_{jij}
+ \beta_{jji}\right)$. In the C$_{3v}$ point group, these relations reduce to
$\bar{\alpha} =\frac{1}{3}(2 \alpha_{yy}+\alpha_{zz})$, $\beta_{\parallel z}
=\frac{3}{5}(2 \beta_{yyz}+\beta_{zzz})$. We also calculate the
hyperpolarizability for hyper-Rayleigh scattering in the VV polarization,
as given by
Cyvin \textit{et al.}\cite{cyvin65} for the static case (where Kleinman
symmetry holds) and the generalization of Bersohn \textit{et
al.}\cite{bersohn66} for the dynamic case. In the static case for C$_{3v}$
symmetry
\begin{equation}
\left[\beta_{\mathrm{HRS}}^{\mathrm{VV}}\right]^2 =
    \frac{8}{35}\beta_{yyy}^2+\frac{1}{7}\beta_{zzz}^2+
    \frac{24}{35}\beta_{yyz}^2+\frac{12}{35}\beta_{yyz}\beta_{zzz} .
\end{equation}
This quantity has only been measured for liquid chloroform;\cite{kaatz98}
measurements are not available for the gas phase.


Our best results for each method generally exhibit a consistent agreement
among themselves and lend confidence to the overall quality of our
calculations compared to earlier work. Achieving this consistency
points to the need for a comprehensive and well balanced
description of all regions of the system: namely,
the outlying regions of the molecule, the short C-H
bond, and the Cl atoms. A key finding is that the local contributions to the
$\beta_{zzz}$ response of the Cl atoms and the C-H bond are of opposite sign and
tend to cancel, thus explaining the relatively slow convergence
of this component with respect to the basis set size. This behavior, together
with the near cancellation of the $\beta_{yyz}$ and $\beta_{zzz}$ components,
leads to
the relatively small value of $\beta_{\parallel}$ of chloroform. By contrast,
the HRS hyperpolarizability converges much more quickly since it is an
incoherent process which is mostly given by a sum of squares of tensor
components that do not cancel.

\section{Methods}
\label{sec:meth}

\subsection{Gaussian-Type Orbitals}
\label{subsec:meth:gtos}

The GTO polarization properties were calculated using finite-field perturbation
theory (FFPT). The electric-field strengths $\mathcal{E}$ used were 0.00, $\pm$0.01 and
$\pm$0.02 au. The different components of the induced dipole moment were fit to a
4th-order polynomial to obtain the polarizability and hyperpolarizability
tensors. Using analytic derivatives available at the Hartree-Fock level, we find
that the properties obtained with the FFPT method agree to 0.1\% or better. The
effects of the basis set size and diffuse quality were studied using Dunning's
double- through quintuple-$\zeta$ correlation-consistent
sets,\cite{dunning89,woon93} with and without
augmentation exponents, and with additional diffuse exponents. We also used
Sadlej's HyPol basis set,\cite{pluta98} which is specifically designed for the
calculation of nonlinear response properties. In this paper, for simplicity,
these basis sets will be referred to as aVDZ and VDZ (for aug-cc-pVDZ and
cc-pVDZ, respectively), aVTZ and VTZ (for aug-cc-pVTZ and cc-pVTZ), etc.
The HyPol set will be abbreviated as HP. 
The basis set
labeled aV5Zs corresponds to the aV5Z set, where the $g$ and $h$ functions were
removed from the C and Cl atoms and the $f$ and $g$ were removed from the H
atom. The d-aV5Zs basis set corresponds to the aV5Zs set augmented with (0.014184,
0.009792, 0.025236)
and (0.017244, 0.012528, 0.036108)
($s$,$p$,$d$) exponents on the C and Cl atoms, respectively and
(0.004968, 0.026784)
($s$,$p$) exponents on the H atom.

Throughout this paper,
unless otherwise specified, we use Hartree atomic units $e=\hbar=m=1$ with
distances in Bohr ($a_0\approx$ 0.529 \AA) and energies in Hartrees $\approx$
27.2 eV.  The effect of electron
correlation on these all-electron calculations was studied with the
LDA,\cite{LDA} PBE,\cite{PBE} B3LYP,\cite{Becke, B3LYP} and BMK\cite{BMK}
exchange-correlation density functionals, as well as with the HF, MP2 and CCSD
methods. Other XC functionals, such as the CAM-B3LYP functional,\cite{CAMB3LYP}
have been used for hyperpolarizability calculations, but we have not included them
since they have not shown a systematic improvement with respect to CCSD.\cite{Hammond}
All GTO calculations were performed with Gaussian 03.\cite{frisch03}

\subsection{Real-Space Grids}
\label{subsec:meth:rsgs}

For the real-space grid calculations we used {\it ab initio} density-functional
theory with a real-space basis, as implemented in
Octopus.\cite{octopus-castro,Marques200360} The polarizability and
hyperpolarizability were calculated by linear response via the Sternheimer
equation and the $2n+1$ theorem.\cite{andrade:184106} This approach, also known
as density-functional perturbation theory, avoids the need for sums over
unoccupied states. The PBE exchange-correlation functional was used for the
ground state, and the adiabatic LDA kernel was used for the linear response. All
calculations used Troullier-Martins norm-conserving
pseudopotentials.\cite{Troullier-Martins}

The molecule was studied as a finite system, with zero boundary conditions for the wavefunction on a
large sphere surrounding the molecule, as described below. Convergence was
tested with respect to the real-space grid spacing and the radius of the
spherical domain. The grid spacing required is determined largely by the
pseudopotential, as it governs the fineness with which spatial variations of the
wavefunctions can be described as well as the accuracy of the finite-difference
evaluation of the kinetic-energy operator. The spacing $\lambda$ can be converted to an
equivalent plane-wave cutoff via $E_c = ({\hbar^2}/{2 m})({2 \pi}/{\lambda})^2$,
where $E_c$ is the cutoff energy for both the charge density
and wavefunctions. The
sphere radius determines the maximum spatial extent of the wavefunctions.

With tight numerical tolerances in solving the Kohn-Sham and Sternheimer
equations, we can achieve a precision of 0.01 au or better in the converged
values of the tensor components of $\beta$. We also did two additional kinds of
calculations. For comparison to the nonlinear experiments, which used incoming
photons of wavelength 1064 nm (energy 1.165 eV), we also performed dynamical
calculations at this frequency via time-dependent density-functional theory
(TDDFT). To compare directly to the results from finite-field perturbation
theory with the other basis sets, we also calculated the dielectric properties
via finite differences using electric-field strengths of $\pm$0.01 and $\pm$0.015 au.

\subsection{Numerical Basis Sets}
\label{subsec:meth:numbas}

The numerical basis set (NBS) calculations were performed with the {\sc
Siesta}\cite{artacho99,soler02}
code and used
Troullier-Martins norm-conserving pseudopotentials.\cite{Troullier-Martins} As
in the GTO calculations described in section \ref{subsec:meth:gtos}, the
polarization properties were obtained using FFPT with electric-field strengths
of 0.00, $\pm$0.01 and $\pm$0.02 au. The NBSs use a generic
linear combination of numerical atomic orbitals (NAOs)
that are forced to be zero beyond a cutoff radius $r_c$.\cite{sankey}
Rather than using a fixed $r_c$ for all atoms, a common
confinement-energy shift is usually enforced, resulting in well-balanced basis
sets.\cite{soler02}
In general,
multiple radial functions
with the same angular dependence are introduced to enhance the flexibility of
the basis set.
This
results in a so-called multiple-$\zeta$ scheme similar to the standard
split-valence approach used for GTOs.\cite{huzinaga85,szabo82} Polarization
functions can also be added using the approach described in Ref.
\onlinecite{soler02}.
Typically, double-$\zeta$ sets with a single polarization function (DZP)
are sufficient in linear-response calculations.
However, we found that a DZP set is insufficient for nonlinear
properties. Instead of performing an optimization of the NBSs, we improved
their flexibility by adding $\zeta$-functions,
and controlling their splitting.\cite{soler02}
By varying the ``norm-splitting'' parameter in {\sc Siesta}, we can control
the flexibility in different regions of the radial functions,
resulting in the cutoff radii and energy
shifts
shown in Tables \ref{tbl:numbas} and \ref{tbl:numbas2}.
\begin{table}[htbp]
\caption{
Parameters used in the definition of the numerical basis sets: energy shifts
$\delta \epsilon$ and cutoff radii $r_c^l(\mathrm{X})$ of the first-$\zeta$ for
angular momentum $l$ and atom X. All values are in atomic units.
}
\label{tbl:numbas}
\begin{ruledtabular}
\begin{tabular}{cdddddd}
\tbhr{Parameter}           &\tbhc{DZP} &\tbhc{QZTPe4} &\tbhc{5Z4Pe5}   &\tbhc{5Z4Pe6}
		    &\tbhc{5Z4Pe7} &\tbhc{5Z4Pe8} \\
\hline
$\delta \epsilon$          & \tbr{$10^{-2}$} & \tbr{$10^{-4}$}& \tbr{$10^{-5}$}& \tbr{$10^{-6}$} & \tbr{$10^{-7}$} & \tbr{$10^{-8}$} \\
$r_c^s(\mathrm{C})$ & 4.088     & 6.574     & 7.832     & 8.875     &10.056     &11.114 \\
$r_c^p(\mathrm{C})$ & 4.870     & 8.655     &10.572     &12.283     &14.271     &15.772 \\
$r_c^s(\mathrm{H})$ & 4.709     & 8.164     & 9.972     &11.586     &13.461     &14.877 \\
$r_c^s(\mathrm{Cl})$& 3.826     & 5.852     & 6.799     & 7.704     & 8.730     & 9.410 \\
$r_c^p(\mathrm{Cl})$& 4.673     & 7.514     & 8.951     &10.400     &11.784     &13.024 \\
\end{tabular}
\end{ruledtabular}
\end{table}
\begin{table}[htbp]
\caption{
Parameters used in the definition of the numerical basis sets: splitting radii
$r_c^l(\mathrm{X})$ associated with a given norm splitting for angular momentum
$l$ and atom X. All values are in atomic units.
}
\label{tbl:numbas2}
\begin{ruledtabular}
\begin{tabular}{cdddddd}
\tbhr{Norm splitting}    &\tbhr{0.0015} &\tbhr{0.0150} &\tbhr{0.1500}   &\tbhr{0.6000}\\
\hline
$r_c^s(\mathrm{C})$   & 6.574     & 5.120     & 3.519     &2.272\\
$r_c^p(\mathrm{C})$   & 8.548     & 6.332     & 3.841     &2.005\\
$r_c^s(\mathrm{H})$   & 8.690     & 6.600     & 4.155     &2.223\\
$r_c^s(\mathrm{Cl})$  & 5.707     & 4.557     & 3.252     &2.292\\
$r_c^p(\mathrm{Cl})$  & 7.328     & 5.566     & 3.639     &2.235\\
\end{tabular}
\end{ruledtabular}
\end{table}
Finally, the NBS calculations used a common (20.0 \AA)$^{3}$ cell and real-space
grid with a plane-wave-equivalent cutoff of 250 Ry for the calculation of the
Hartree and exchange-correlation potentials. This corresponds to a real-space
mesh spacing of about 0.1 \AA.

\subsection{Linear and Nonlinear Response Densities}
\label{subsec:meth:prop}

The origin of the slow convergence of the hyperpolarizability with respect to
the quality of the basis set is difficult to understand by studying only the
total quantities. A more informative analysis can be obtained from the
spatial distribution of the dielectric properties. Thus, we have calculated the
linear and nonlinear response densities, as well as their associated
properties. Here we will focus on the response densities induced by an electric
field in the $z$-direction. The first-order density is defined as
\begin{equation}
  \rho_z^{(1)}\left(r \right) = \frac{\partial \rho}{\partial \mathcal{E}_z},
\end{equation}
and the linear polarizability $\alpha_{zz}\left(r \right)$ as
\begin{equation}
  \alpha_{zz}\left(r \right) = \rho_z^{(1)}\left( r \right) z .
\end{equation}
The second-order response density and associated hyperpolarizability are defined
similarly:
\begin{equation}
  \rho_{zz}^{(2)}\left(r \right) =
  \frac{\partial^2 \rho}{\partial \mathcal{E}_z^2},
\end{equation}
\begin{equation}
   \beta_{zzz} \left( r \right) = \rho_{zz}^{(2)}\left( r \right) z .
\end{equation}
These response densities are all calculated using finite differences.
For the real-space grids, our Sternheimer approach provides 
only the linear response density and polarizability density, but not the
nonlinear response and hyperpolarizability densities.

Unlike the total properties, the spatial distributions of polarizabilities and
hyperpolarizabilities as defined above depend on the origin of coordinates.
Throughout this work we chose a center-of-mass reference for the spatial
distributions. To understand the role that different regions of the molecule
play in the total properties, we have devised a partitioning scheme for the
spatial distribution corresponding to the spaces occupied roughly by
the Cl atoms and C-H bond. That is, we divide space into two regions 
 by constructing three planes, each orthogonal to one of the C-Cl bonds,
and passing through a point located 40\% along the C-Cl bond from the C atom,
which corresponds approximately to the density minimum along the C-Cl bond.
The first region (``CH'') consists of all the space above the three planes,
and contains the C-H bond, while the second (``Cl'') covers the remainder 
of the space, including the three Cl atoms.
We have integrated the various densities in each region numerically to find
its contribution to the total.






\section{Results and Discussion}
\label{sec:res}

\subsection{Structure}
\label{subsec:struc}

Although all the results presented in later sections were obtained for the
experimental geometry determined by Colmont \textit{et
al.},\cite{Colmont1998158} here we briefly discuss the effect of the structural
parameters on the dielectric properties. We compared properties obtained for
experimental structures\cite{Colmont1998158,jen:2525} with theoretical
structures optimized using the PBE functional, one obtained with the aVQZ basis in
Gaussian03,\cite{davidson06} and the other with a real-space grid in Octopus.
The parameters for each structure are listed in Table \ref{tab:Geometries}. The
linear and nonlinear properties for each structure were calculated with the
Sternheimer method in Octopus, using a radius of 17 $a_0$ and a spacing of 0.25
$a_0$ and the results are summarized in Table \ref{tab:Structure properties}.
Our calculations show that the dipole moment and polarizability are not affected
much, but the hyperpolarizability varies significantly with structure.
Individual tensor components of the hyperpolarizability do not change by more
than $\sim$30\%, but since $\beta_{\parallel}$ is a sum of large positive and
negative components, it can change sign, and change by orders of magnitude
depending on the structure. Note, however, that all geometries give results
consistent with the large relative error bar on the gas-phase experimental value
of $\beta_{\parallel} = 1 \pm 4$ au.\cite{kaatz98}

\begin{table}

\caption{Structural parameters used in the study of the variation of the
dielectric properties of CHCl$_3$ with structure. PBE/aVQZ and PBE/RS refer to
PBE-optimized structures using the aVQZ GTO in Gaussian and a real-space grid in
Octopus, respectively.
Bond lengths are in
\AA\ and angles in degrees.
The experimental structure from Ref. \onlinecite{Colmont1998158} was used for
all our subsequent calculations.}
\label{tab:Geometries}
\begin{ruledtabular}
\begin{tabular}{lllll}
\tbhl{Source}   & \tbhc{$r$(C-H)} & \tbhc{$r$(C-Cl)} & \tbhc{\angle HCCl} \\
\hline
Expt.\cite{jen:2525}       & 1.100 $\pm$ 0.004 & 1.758 $\pm$ 0.001 & 107.6  $\pm$ 0.2  \\
Expt.\cite{Colmont1998158}  & 1.080 $\pm$ 0.002 & 1.760 $\pm$ 0.002 & 108.23 $\pm$ 0.02 \\
PBE/aVQZ\cite{davidson06}  & 1.090             & 1.779             & 107.7  \\
PBE/RS               & 1.084             & 1.769             & 107.6  \\
\end{tabular}
\end{ruledtabular}
\end{table}

\begin{table}
\caption{Dielectric properties of various structures for
CHCl$_3$ described in Table \ref{tab:Geometries}, as calculated
by DFT on a real-space grid with radius 17 $a_0$ and spacing 0.25 $a_0$,
compared with the experimental values of the dipole moment and
the electronic contribution to the polarizability.
PBE/aVQZ and PBE/RS refer to the structures described in Table
\ref{tab:Geometries}.
All values are in atomic units (au).}
\label{tab:Structure properties}
\begin{ruledtabular}
\begin{tabular}{lddddddddd}
\tbhc{Structure} &\tbhc{$\mu_z$} &\tbhc{$\alpha_{yy}$}  &\tbhc{$\alpha_{zz}$} &\tbhc{$\beta_{yyy}$} &\tbhc{$\beta_{yyz}$} &\tbhc{$\beta_{zzz}$} &\tbhc{$\bar{\alpha}$} &\tbhc{$\beta_{\parallel}$} &\tbhc{$\beta_{\mathrm{HRS}}^{\mathrm{VV}}$}\\    
\hline
Expt.\cite{jen:2525}       & 0.395   & 66.14 & 47.22 & 27.09 &  -14.41   & 28.47  &  59.83   & -0.21 & 16.89 \\
Expt.\cite{Colmont1998158} & 0.399   & 66.02 & 47.00 & 27.12 &  -16.36   & 26.92  &  59.68   & -3.49 & 17.44 \\
PBE/aVQZ\cite{davidson06}    & 0.401   & 67.17 & 47.35 & 27.23 &  -14.11   & 27.76  &  60.57   & -0.27 & 16.79 \\
PBE/RS                   & 0.397   & 66.66 & 47.12 & 27.29 &  -14.26   & 28.92  &  60.15   &  0.24 & 16.96 \\

\hline
\tbhl{Expt.} & \tbr{0.409$\pm$0.008}\cite{reinhart70} & \tbr{61$\pm$5}\cite{Stuart_Volkmann} & \tbr{45$\pm$3}\cite{Stuart_Volkmann} &  &  &  & \tbr{56$\pm$4}\cite{Stuart_Volkmann} & \tbr{1$\pm$4}\cite{kaatz98} & \\
\end{tabular}
\end{ruledtabular}
\end{table}

\subsection{Gaussian-Type Orbitals}
\label{subsec:gto}

Table \ref{tab:BS_Eff} shows the effect of the basis set size on the dielectric
properties calculated with the PBE functional and GTOs. The results from this
work agree well with those reported by Davidson \textit{et al.}\cite{davidson06}
In the case of the dipole moment, we also find agreement to within 1\% of the
experimental value of 0.409 $\pm$ 0.008 au\cite{reinhart70} for every basis set
except VDZ. (The quality of the basis set roughly increases down the table.) The polarizabilities agree well with the experimentally measured
values at 546 nm (2.27 eV);\cite{Stuart_Volkmann} these quantities are optical
polarizabilities which contain only an electronic contribution and have minimal
dispersion. Indeed, our real-space TDDFT calculations (Section \ref{subsec:rsg})
at 532 nm (2.24 eV) give $\alpha_{yy}$ = 68.827 au and $\alpha_{zz}$ = 48.405
au, a small increase from the static and 1064 nm results, and basically
consistent with the experimental values. We can also compare to a measurement of
the static isotropically averaged polarizability of 64 $\pm$ 3
au.\cite{miller:4858} To compare with our result for the average electronic
polarizability, we subtract the estimated vibrational contribution of 4.5 au
calculated from experimental data (no error bar provided),\cite{bishop82}
yielding 60 au, which agrees with the predicted values within 0.4\%. To verify
this comparison, we have also computed the vibrational component of the
polarizability from first principles, obtaining a value of 6.3 au, in reasonable
agreement with the above estimate. This value was obtained by first doing a
standard Gaussian 03\cite{frisch03} calculation of vibrational-frequencies at
the PBE/aug-cc-pVTZ level, using a molecular structure optimized at the same
level. The polarizability is then  calculated as described in Ref.
\onlinecite{kirtman:4125}.

\begin{table}
\caption{
Effect of the GTO basis set quality on the components of the dielectric
properties of CHCl$_3$ calculated with the PBE functional. All values are in atomic
units.
}
\label{tab:BS_Eff}
\begin{ruledtabular}
\begin{tabular}{lddddddddd}
\tbhl{Basis Set}  & \tbhc{$\mu_{z}$} & \tbhc{$\alpha_{yy}$} & \tbhc{$\alpha_{zz}$} & \tbhc{$\beta_{yyy}$} & \tbhc{$\beta_{yyz}$} & \tbhc{$\beta_{zzz}$} & \tbhc{$\bar{\alpha}$} & \tbhc{$\beta_{\parallel}$} & \tbhc{$\beta_{\mathrm{HRS}}^{\mathrm{VV}}$} \\
\hline
VDZ	&  0.426  &    46.36	&   25.82    &    19.64    &    -8.43    &	-44.25   &   39.51   &   -36.66  & 23.33 \\
VTZ	&  0.417  &    54.57	&   34.08    &     0.52    &    -3.35    &	-37.20   &   47.74   &   -26.33  & 15.75 \\
VQZ	&  0.407  &    59.92	&   39.76    &    -5.83    &    -2.88    &	-25.57   &   53.20   &   -18.80  & 10.90 \\
V5Z	&  0.405  &    61.65	&   41.93    &    -6.12    &    -5.70    &	-25.85   &   55.08   &   -22.35  & 12.64 \\
aVDZ	&  0.412  &    63.29	&   44.35    &     8.53    &   -11.35    &	  2.62   &   56.98   &   -12.05  &  9.78 \\
aVTZ	&  0.408  &    64.91	&   46.03    &    14.67    &   -11.81    &	 11.72   &   58.62   &    -7.14  & 10.82 \\
aVQZ	&  0.406  &    65.50	&   46.61    &    21.15    &   -13.81    &	 19.52   &   59.21   &    -4.86  & 13.96 \\
aV5Z	&  0.404  &    65.69	&   46.71    &    22.02    &   -14.26    &	 18.65   &   59.36   &    -5.92  & 14.45 \\
aV5Zs	&  0.404  &    65.67	&   46.70    &    21.78    &   -14.01    &	 17.92   &   59.34   &    -6.07  & 14.24 \\
d-aV5Zs	&  0.404  &    65.70	&   46.79    &    27.35    &   -15.31    &	 22.27   &   59.40   &    -5.01  & 16.90 \\
HP	&  0.405  &    65.51	&   46.60    &    27.64    &   -15.54    &	 22.93   &   59.21   &    -4.89  & 17.12 \\
\hline
\tbhl{Expt.} & \tbr{0.409$\pm$0.008}\cite{reinhart70} & \tbr{61$\pm$5}\cite{Stuart_Volkmann} & \tbr{45$\pm$3}\cite{Stuart_Volkmann} &  &  &  & \tbr{56$\pm$4}\cite{Stuart_Volkmann} & \tbr{1$\pm$4}\cite{kaatz98} & \\
\end{tabular}
\end{ruledtabular}
\end{table}

For the hyperpolarizability, Davidson \textit{et al.} also obtain good agreement
between theory and experiment provided the values of $\beta_{\parallel}$
calculated with the aVDZ, aVTZ and aVQZ correlation-consistent basis sets are
smoothly extrapolated to the complete basis set limit.\cite{peterson93} If the
same extrapolation scheme is applied to our results for $\beta_{\parallel}$
given in Table \ref{tab:BS_Eff}, however, we obtain a value of -2.89 au, which
is barely within the error of the experimental value of 1 $\pm$ 4 au. This
extrapolated value also differs from the value of 0.35 au obtained by Davidson
\textit{et al.} The difference can be attributed to the different geometries
used: In this work we used the experimentally determined structure, while
Davidson \textit{et al.} used the theoretical structures obtained with the aVDZ,
aVTZ and aVQZ basis sets. When we use the aVTZ-optimized geometry in our
calculations instead of the experimental one, we obtain an extrapolated value of
0.6 au, which is consistent with the Davidson \textit{et al.} value. Our
calculations also included the aV5Z basis set, the next basis set in the
correlation-consistent series. When this set is included in the extrapolation,
our predicted value is lowered to -5.29 au. The sudden reduction of the
estimated complete basis set value upon inclusion of the aV5Z set indicates that
the convergence of $\beta_{\parallel}$ is not smoothly monotonic. Therefore the
basis set sequence aVDZ-aV5Z is not well adapted to the extrapolation of this
property. This stems from the fact that although the individual components
$\beta$ rise in a reasonably monotonic way, small deviations in their
progression and their near cancellation lead to sudden changes in
$\beta_{\parallel}$.

\begin{table}
\caption{
Effect of the exchange-correlation (XC) treatment on the components of the
dielectric properties of CHCl$_3$ calculated with the HP GTO basis set. All
values are in atomic units.
}
\label{tab:BS_Eff2}
\begin{ruledtabular}
\begin{tabular}{lddddddddd}
\tbhl{XC}    & \tbhc{$\mu_{z}$} & \tbhc{$\alpha_{yy}$} & \tbhc{$\alpha_{zz}$} & \tbhc{$\beta_{yyy}$} & \tbhc{$\beta_{yyz}$} & \tbhc{$\beta_{zzz}$} & \tbhc{$\bar{\alpha}$} & \tbhc{$\beta_{\parallel}$} & \tbhc{$\beta_{\mathrm{HRS}}^{\mathrm{VV}}$} \\
\hline
LDA   &  0.414  &    65.95   &   47.03  &	28.89  &  -16.21  &	 22.56  &   59.64  &   -5.91  & 17.83 \\
PBE   &  0.405  &    65.51   &   46.60  &	27.64  &  -15.54  &	 22.93  &   59.21  &   -4.89  & 17.12 \\
BMK   &  0.453  &    62.13   &   44.77  &	22.56  &  -11.58  &	 18.63  &   56.35  &   -2.72  & 13.56 \\
B3LYP &  0.423  &    63.60   &   45.60  &	21.90  &  -12.99  &	 18.98  &   57.60  &   -4.21  & 13.87 \\
HF    &  0.476  &    58.98   &   42.76  &	13.57  &   -7.58  &	 13.02  &   53.57  &   -1.28  &  8.48 \\
MP2   &  0.424  &    62.13   &   45.18  &	15.03  &   -9.75  &	 16.47  &   56.48  &   -1.82  & 10.03 \\
CCSD  &  0.425  &    61.54   &   44.88  &	16.51  &  -10.31  &	 16.81  &   55.99  &   -2.29  & 10.78 \\
\hline
\tbhl{Expt.} & \tbr{0.409$\pm$0.008}\cite{reinhart70} & \tbr{61$\pm$5}\cite{Stuart_Volkmann} & \tbr{45$\pm$3}\cite{Stuart_Volkmann} &  &  &  & \tbr{56$\pm$4}\cite{Stuart_Volkmann} & \tbr{1$\pm$4}\cite{kaatz98} & \\
\end{tabular}
\end{ruledtabular}
\end{table}


To better understand the convergence of the nonlinear properties with respect to
the degree of polarization and diffuseness of the basis sets, we also performed
calculations using simplified and enhanced versions of the
correlation-consistent basis sets. The results for the non-augmented sets
(labeled VDZ-V5Z in Table \ref{tab:BS_Eff}) indicate, as is widely known, that
the diffuse exponents are important for the polarizability and crucial for the
hyperpolarizability. Even the very large V5Z basis set yields
$\beta_{\parallel}$ values that are too low. These values are also less well
converged than the much smaller aVDZ augmented set. The polarization functions
with high angular momentum (\textit{i.e.} larger than $d$ and $f$ for the
hydrogen and heavy atoms, respectively) play a very small role, as seen from the
similarity between the results obtained with the aV5Z set, and the aV5Zs set,
where such functions were removed. The results in Table \ref{tab:BS_Eff} show
that this simplification has a negligible effect on the dipole moment and linear
polarizability. It also has a fairly small effect ($\sim$3\%) on the
hyperpolarizability.

The hyperpolarizability values can be further improved by enhancing the
diffuseness of the basis set beyond that of the standard Dunning
correlation-consistent sets. When the aV5Zs set is extended with a single $d$-function
with exponent 0.036108 localized on the Cl atoms, the value of
$\beta_{zzz}$ was increased by almost 20\%. Further extending the basis set with
diffuse functions in the C and H atoms, resulting in the d-aV5Zs set, increases
$\beta_{zzz}$ slightly. The d-aV5Zs basis set provides the most saturated GTO
results obtained in this work. We note that, from the point of view
of augmentation, the basis set d-aV5Zs is equivalent to a d-aug-cc-pV5Z set.
That is, the diffuse exponents used are the same as those in the Dunning set.
The main difference between the two sets is in the high-angular-momentum
exponents, which play a very small role as discussed below.
We also attempted to perform the calculations with the t-aug-cc-pV5Z set;
however, Gaussian 03 calculations have convergence problems due to 
the extremely diffuse exponents.

We have also investigated the effect of the tight d-functions on Cl by
carrying out calculations with the cc-pV(5+d)Z  basis set for Cl
augmented with diffuse functions and paired with the d-aV5Zs set
from our original calculations for the C and H atoms. However,
we find that the tight $d$ functions change the individual components by
only about 0.3\% and $\beta_{\parallel}$ by only about 2\%.
Finally, we find that the HyPol basis set, which
is approximately equivalent in size to the aVTZ set, yields results that are
equivalent to the much larger d-aV5Zs set. This basis set was designed with the
explicit purpose of efficiently calculating nonlinear properties. Thus it is not
surprising that it provides converged results for a smaller size.

The large variations in $\beta_{\parallel}$ are due largely to changes in
$\beta_{zzz}$, for which completing the basis set produces a change of sign and
an absolute change of nearly two orders of magnitude. This effect translates
into a change of $\beta_{\parallel}$ of nearly an order of magnitude! The same
behavior is observed as we complete the numerical basis sets (see Table
\ref{tbl:staticchcl3eshifts}) but not in the case of the real-space calculations
(see Table \ref{tab:Results}). Then the two basis set approaches (GTO and NBS)
show similar convergence features as we increase the quality of the basis set.

The variation in the results with the quality of the exchange-correlation
treatment is shown in Table \ref{tab:BS_Eff2}, in order of increasing level
of exchange-correlation accuracy. Note first that the {\it linear} properties
are rather insensitive to the treatment of exchange and correlation. Also the
higher-quality \textit{ab initio} correlation treatments (MP2 and CCSD) are more
or less consistent, while the DFT functionals LDA, PBE, BMK, and B3LYP vary
considerably for the nonlinear properties. In the case of the higher-level
methods, 
the properties follow the usual pattern of decreasing the dipole moment
and enhancing the polarizability when the treatment of correlation is improved.
Compared to the CCSD values, 
the components of the hyperpolarizability
vary by as much as 75\%. Judging by the values of $\beta_{\parallel}$,
the predicted value increases with improvements in the treatment of 
correlation. Also the variation among the values can be regarded as
an estimate of the error in the results due to approximations in the
treatment of exchange and correlation. 

\subsection{Real-Space Grids}
\label{subsec:rsg}

Convergence of the dipole moment, polarizability, and hyperpolarizability is
illustrated in Table \ref{tab:Results}. The total energy was well converged for
a spacing of 0.35 $a_0$ (equivalent plane-wave cutoff = 20 Ry) and a sphere
radius of 12 $a_0$. The dipole moment was also well converged with that basis.
However, to converge $\beta_{\parallel}$ to better than 0.01 au, a spacing of
0.25 $a_0$  (equivalent plane-wave cutoff = 40 Ry) and a sphere radius of 22
$a_0$ was required. The convergence of the tensor components of $\beta$ is
similar to that of $\beta_{\parallel}$ in absolute terms, \textit{i.e.} they are
also converged to 0.01 au or better with these parameters. Generally, the
magnitude of $\beta_{\parallel}$ declines with smaller spacing and larger
radius, as the cancellation between the tensor components becomes closer.

Finite-difference calculations were done with the converged grid spacing of 0.25
$a_0$, and a sphere radius of 22.\ $a_0$, for comparison to the Sternheimer
calculation with the same grid parameters (Table \ref{tab:Best_Res}). The
differences between the linear-response and finite-difference calculations are
small, allowing a direct comparison between the results with different basis sets.
The use of the LDA kernel in the linear-response results gives a small discrepancy
compared to the purely PBE finite-difference results.
Fields of 0.015 au rather than 0.02 au as for the other basis sets were used
because 0.02 au was out of the linear regime in the real-space calculation.
The linear response density $\rho_z^{(1)}(r)$ and polarizability
density $\alpha_{zz}(r)$ are virtually identical between the finite-difference
and linear-response calculations.

Calculations at 1064 nm with the same grid parameters show increases of about
1\% in the polarizability, and 10-20\% in the hyperpolarizability, as compared
to the static case. We find a small violation of Kleinman symmetry here, in that
$\beta_{yyz}$ = -18.945 au whereas $\beta_{zyy}$ = -19.448 au.

\begin{table}
\caption{
Effect of the real-space-grid quality (radius $R$ and spacing $\lambda$)
on the components of the dielectric
properties of CHCl$_3$ calculated with the PBE functional and LDA kernel. All values are in atomic
units.
}
\label{tab:Results}
\begin{ruledtabular}
\begin{tabular}{cddddddddddd}
\tbhc{$R$} &\tbhc{$\lambda$} &\tbhc{$\mu_z$}
&\tbhc{$\alpha_{yy}$}  &\tbhc{$\alpha_{zz}$} &\tbhc{$\beta_{yyy}$} &\tbhc{$\beta_{yyz}$} &\tbhc{$\beta_{zzz}$} & \tbhc{$\bar{\alpha}$} &\tbhc{$\beta_{\parallel z}$} & \tbhc{$\beta_{\mathrm{HRS}}^{\mathrm{VV}}$} \\    
\hline                                                                                                                    
  12	&   0.25  &  0.398 & 65.921 & 46.924 & 27.975 & -17.232 & 22.975 & 59.589 & -6.921  & 17.106 \\
  15	&   0.25  &  0.399 & 66.019 & 46.993 & 27.159 & -16.401 & 26.758 & 59.677 & -3.629  & 17.461 \\
  17    &   0.25  &  0.399 & 66.022 & 46.995 & 27.123 & -16.363 & 26.921 & 59.680 & -3.485  & 17.443 \\
  20	&   0.25  &  0.399 & 66.023 & 46.995 & 27.119 & -16.358 & 26.940 & 59.680 & -3.469  & 17.441 \\
  22	&   0.25  &  0.399 & 66.023 & 46.995 & 27.119 & -16.358 & 26.940 & 59.680 & -3.468  & 17.441 \\
\hline                
  17    &   0.35  &  0.397 & 66.032 & 47.002 & 27.181 & -16.233 & 26.921 & 59.689 & -3.351  & 17.415 \\
  17	&   0.30  &  0.399 & 66.029 & 46.989 & 27.168 & -16.357 & 26.893 & 59.683 & -3.492  & 17.455 \\
  17    &   0.25  &  0.399 & 66.022 & 46.995 & 27.123 & -16.363 & 26.921 & 59.680 & -3.485  & 17.443 \\
  17    &   0.20  &  0.398 & 66.021 & 46.993 & 27.091 & -16.355 & 26.903 & 59.678 & -3.488  & 17.427 \\
\hline                
\tbhl{Expt.} & & \tbr{0.409$\pm$0.008}\cite{reinhart70} & \tbr{61$\pm$5}\cite{Stuart_Volkmann} & \tbr{45$\pm$3}\cite{Stuart_Volkmann} &  &  &  & \tbr{56$\pm$4}\cite{Stuart_Volkmann} & \tbr{1$\pm$4}\cite{kaatz98} & \\
\end{tabular}
\end{ruledtabular}
\end{table}

\subsection{Numerical Basis Sets}

Table \ref{tbl:staticchcl3eshifts} shows the variation of the dielectric
properties as a function of the quality of the numerical basis set; this is
determined by the ``energy shift'' parameter $\delta \epsilon$ (which is a
measure of the spatial extent of the basis) and the number of atomic orbitals.\cite{artacho99}
The default DZP basis set, using the default energy shift of 0.01 au, produces
results that are clearly inadequate, being unable to reproduce even the dipole
moment. We have found for CHCl$_3$, that $\delta \epsilon$ must be at least
$0.5\times 10^{-6}$ au   to obtain a value of $\beta_\parallel$ that approaches
those obtained with GTOs and real-space grids. Numerical basis set convergence
is apparently achieved with $\delta \epsilon = 1\times 10^{-8}$ au. The
corresponding cutoff radii are 2 to 3 times larger than those produced by the
default value of $\delta \epsilon$, clearly matching the need for very diffuse
GTOs and large confining spheres for the real-space grids. Moreover, both the
valence and polarization parts of the numerical basis set have to be
sufficiently flexible to represent the response properties accurately. For
instance, at least five valence and four polarization atomic orbitals are needed
to obtain accurate results, as is also the case for the GTOs (\textit{e.g.} the
d-aV5Zs set has five $sp$ valence functions plus four $d$- and three
$f$-polarization functions). The best value of $\beta_\parallel$ is -5.07 au,
obtained with the 5Z4Pe8 numerical basis set. This is in good agreement with the
value of -5.01 au obtained with the d-aV5Zs GTO set. The close agreement with the
GTO results seems to be fortuitous since the components of $\beta$ differ by
about 6\%.


\begin{table}[htbp]
\caption{
Effect of the numerical basis set quality on the components of the dielectric
properties of CHCl$_3$ calculated with the PBE functional. All values are in atomic
units. See Table \ref{tbl:numbas} for a detailed description of each basis set.
Note that DZP and DZPe4 are both double-$\zeta$ basis with the same split radius
except the first-$\zeta$'s cut-off radius is the same as QZTPe4 for DZPe4.
}
\label{tbl:staticchcl3eshifts}
\begin{ruledtabular}
\begin{tabular}{lddddddddd}
\tbhl{Basis Set} & \tbhc{$\mu_{z}$} & \tbhc{$\alpha_{yy}$} & \tbhc{$\alpha_{zz}$} & \tbhc{$\beta_{yyy}$} & \tbhc{$\beta_{yyz}$} & \tbhc{$\beta_{zzz}$} & \tbhc{$\bar{\alpha}$} & \tbhc{$\beta_{\parallel}$} & \tbhc{$\beta_{\mathrm{HRS}}^{\mathrm{VV}}$} \\
\hline
DZP    & 0.240 & 47.30 & 29.99  &  11.68 &  -5.35 & -29.15 & 41.53 & -23.91  & 15.02  \\
DZPe4  & 0.361 & 56.45 & 37.62  &   6.05 &  -9.00 & -39.98 & 50.17 & -34.79  & 20.39  \\
QZTPe4 & 0.392 & 63.71 & 44.69  &  12.18 & -10.52 &  -9.90 & 57.37 & -18.57  & 12.63  \\
5Z4Pe5 & 0.397 & 65.14 & 46.02  &  21.93 & -13.70 &  14.16 & 58.77 &  -7.96  & 14.17  \\
5Z4Pe6 & 0.398 & 65.41 & 46.23  &  23.71 & -14.77 &  19.51 & 59.02 &  -6.02  & 15.29  \\
5Z4Pe7 & 0.398 & 65.45 & 46.28  &  24.48 & -14.93 &  20.18 & 59.06 &  -5.22  & 15.64  \\
5Z4Pe8 & 0.398 & 65.45 & 46.28  &  24.54 & -14.90 &  21.37 & 59.06 &  -5.07  & 15.68  \\
\hline
\tbhl{Expt.} & \tbr{0.409$\pm$0.008}\cite{reinhart70} & \tbr{61$\pm$5}\cite{Stuart_Volkmann} & \tbr{45$\pm$3}\cite{Stuart_Volkmann} &  &  &  & \tbr{56$\pm$4}\cite{Stuart_Volkmann} & \tbr{1$\pm$4}\cite{kaatz98} & \\
\end{tabular}
\end{ruledtabular}
\end{table}


\subsection{Linear and Nonlinear Response Densities}
\label{subsec:prop}

\begin{figure}[t]
\includegraphics[scale=0.60]{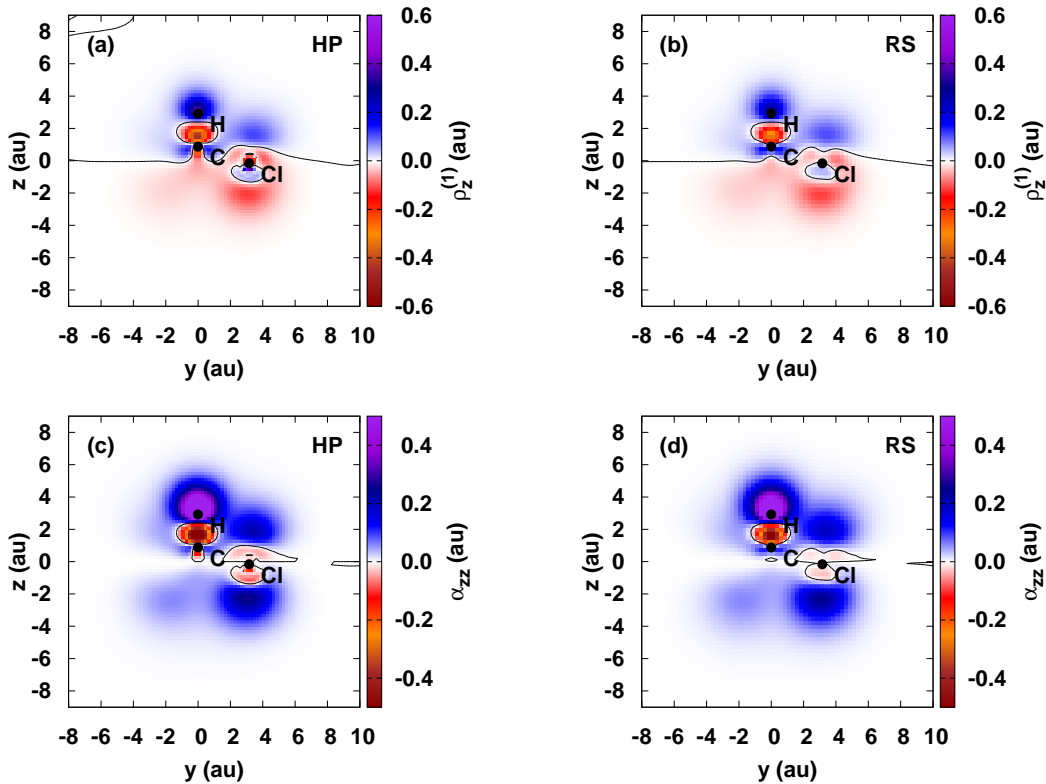}
\caption{\label{fig:rho1-pl}
Linear response density $\rho_z^{(1)}(r)$ (a-b) and polarizability density
$\alpha_{zz}(r)$ (c-d) on one of the HCCl planes of the molecule calculated
with a GTO basis set (HP) and a real-space grid (RS) using the PBE functional.
The positions of the nuclei are indicated with black dots, and the black lines
are isolines. All quantities are
in atomic units. Note that the linear response density is quite similar for both
methods.
}
\end{figure}

The origin of the slow convergence of the response properties is difficult to
determine just by analyzing their total values for different basis sets. We have
found that the differences and similarities between those values can be
visualized by computing the real-space distribution of the linear and nonlinear
response densities, as well as the associated polarizability and
hyperpolarizability densities. For example, Figure \ref{fig:rho1-pl} shows the
linear response density $\rho^{(1)}_z\left(r\right)$ calculated with the PBE
functional for both the HP GTO basis set and a real-space grid. Clearly, the
nearly identical plots confirm that the linear response is equally well
represented by both basis sets. Also shown in Figure \ref{fig:rho1-pl}, the
polarizability density $\alpha_{zz}\left(r\right)$ reveals the spatial
contributions to the total polarizability. For the most part, this property is
localized to within $\sim$ 6 au of the center of the molecule, explaining its
rapid convergence with respect to the diffuseness of the basis set. Our
partitioning scheme for $\alpha_{zz}(r)$ (Table \ref{tab:Den_Part}) shows that
most of the positive contribution to $\alpha_{zz}$ arises from the Cl atoms, in
accord with the larger polarizability of the Cl atom. The contribution from the
CH bond is significantly smaller and, as can be clearly seen in Figure
\ref{fig:rho1-pl}, is the result of counteracting contributions: positive from
the H atom and negative from the C-H bond region. Our partitioning also shows
that the deficiencies of the GTOs are due mostly to a poorer representation of
the Cl atoms.

\begin{figure}[t]
\includegraphics[scale=0.60]{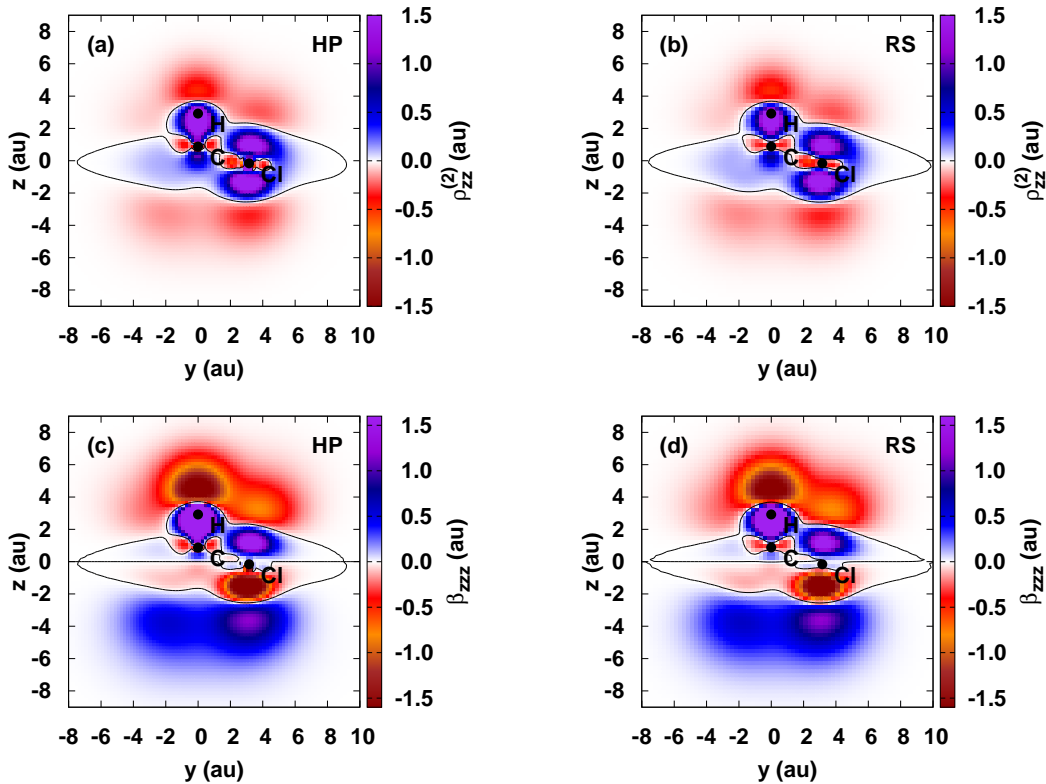}
\caption{\label{fig:rho2-hy}
Nonlinear response density $\rho_{zz}^{(2)}(r)$ (a-b) and hyperpolarizability
density $\beta_{zzz}(r)$ (c-d) on one of the HCCl planes of the molecule
calculated with a GTO basis set (HP) and a real-space grid (RS) using the PBE
functional. The positions of the nuclei are indicated with black dots,
and the black lines are isolines. All
quantities are in atomic units. The nonlinear densities extend much further into
space than the linear densities. The agreement between the real-space and GTO
methods is nevertheless quite good. The contributions to the hyperpolarizability
from the Cl atoms and the CH bond are of opposite sign and, as indicated by the
nonlinear response density, have contributions that extend even further into
space.
}
\end{figure}

The spatial distributions of the nonlinear response density $\rho_{zz}^{(2)}(r)$
and hyperpolarizability density $\beta_{zzz}(r)$, shown in Figure
\ref{fig:rho2-hy}, are also very similar for both the HP GTO set and the
real-space grid. The hyperpolarizability density, however, is more delocalized
than the polarizability, extending up to $\sim$ 8 au from the center, thus
stressing the importance of the diffuse functions in calculations of nonlinear
properties. The spatial distribution is also much more complex, with several
regions of counteracting contributions. The decomposition shown in Table
\ref{tab:Den_Part} significantly simplifies the analysis of the densities. It
shows that the overall contribution from the C-H bond is negative. This
contribution also varies very little with respect to the quality of the basis
set. The contribution from the Cl atoms, on the other hand, is positive and
varies significantly with the basis set used. For instance, the value obtained
for the aVDZ set is almost 30\% lower than the aV5Z set. Even the aV5Z basis set does
not provide converged results, since the inclusion of extra diffuse exponents
(d-aV5Zs set) results in a further increase of 5\%. This change is small for the
individual Cl contribution, but changes the total $\beta_{zzz}$ component by
16\%. Finally, it can again be seen that, although significantly smaller, the HP
set provides results that are almost identical to those obtained with the d-aV5Zs
one.
The plots shown in Figure \ref{fig:rho2-hy} indicate that the CH bond is well
saturated with respect to the extent of the diffuse functions. This observation
was confirmed by using the d-aV5Zs basis set, which also includes diffuse functions
on both the C and H atoms and only slightly improves the results.

\begin{table}
\caption{
Partitioning of the linear and nonlinear response properties calculated
numerically with GTOs and real-space grids (RS) using the PBE functional
together with the numerical sum over the CH and Cl$_3$ partitions.
For comparison the ``Analytic'' results are given for the integral over 
all space (without partitioning) from Table V.
}
\label{tab:Den_Part}
\begin{ruledtabular}
\begin{tabular}{lldddd}
\tbhl{Property}&\tbhl{Basis Set}  & \tbhr{CH} & \tbhr{Cl$_{3}$} & \tbhr{CH +
Cl$_{3}$} & \tbhr{Analytic} \\
\hline
$\alpha_{zz}$     & aVDZ    &  8.91	 &35.42  & 44.33  & 44.35\\
                  & aV5Z    &  8.96      &37.71  & 46.67  & 46.71\\
                  & d-aV5Zs &  8.96      &37.79  & 46.75  & 46.79\\
                  & HP      &  8.92      &37.67  & 46.59  & 46.60\\
                  & RS      &  8.85      &38.02  & 46.87  & 46.87\\
\hline
$\beta_{zzz}$    &  aVDZ    & -43.82  &  46.46 &   2.63 &   2.62\\
                 &  aV5Z    & -44.48  &  63.08 &  18.61 &  18.65\\
                 &  d-aV5Zs & -44.31  &  66.54 &  22.24 &  22.27\\
                 &  HP      & -44.55  &  67.49 &  22.94 &  22.93\\
                 &  RS      & -43.41  &  67.12 &  23.71 &  23.89\\
\end{tabular}
\end{ruledtabular}
\end{table}

\section{Conclusions}
\label{sec:concl}

We have carried out calculations of the static hyperpolarizability of the
gas-phase CHCl$_3$ molecule, using three different kinds of basis sets:
Gaussian-type orbitals, numerical basis sets, and real-space grids.  We find
that all of these methods can yield quantitatively similar
results provided sufficiently large,
diffuse basis sets are included in the calculations. In particular   diffuse
functions are important to obtain accurate results for the polarizability and
are crucial for the hyperpolarizability. For GTOs, the standard versions of the
augmented Dunning basis set are not adequate to converge the $\beta_{zzz}$
component. However, convergence can be achieved by increasing the diffuse $d$
space on the Cl atoms. Other diffuse functions play smaller role.  The
overall consistency among the results gives confidence to their reliability
and overall accuracy.  Based on the size of the basis sets and degree of
convergence, the LR real-space values in Table X are likely the most reliable.
However, the variation among our results also provides a gauge of their overall
theoretical accuracy.

Although the treatment here has been restricted to chloroform, many of
the results are more generally applicable. For example,
the spatial distributions provided by the linear and nonlinear response
densities provides a good visualization tool to understand the basis set
requirements for the simulation of linear and nonlinear response.
A key finding for chloroform is that the local contributions near the
Cl atoms and the CH bond are of opposite sign and tend to
cancel, thus explaining the overall weakness of the hyperpolarizability.
The molecule's response is quite extended in space and so real-space
grids on a large domain, as well as very extended local orbitals, are required
to describe it properly. The frequency-dependence of the polarizability and
hyperpolarizability is small, as verified by our time-dependent calculations,
and so dispersion is not very important in comparing static theoretical results
to experimental measurements.

The discrepancy between the experimentally determined linear polarizability and
our theoretical results is essentially eliminated when the vibrational component
is taken into account. Our results for the hyperpolarizability for all
three basis sets are all consistent with each other.  Given the error bars
in the experimental result our PBE hyperpolarizability results are smaller,
though essentially consistent with the experimental measurements for
$\beta_{\parallel}$, even without taking into
account vibrational contributions. With better treatments of
exchange and correlation (Table VI) the agreement is expected to be further
improved.  Experimental results indicate that the
vibrational contribution is small for the hyperpolarizability: differences in
the hyperpolarizability of isotopically substituted molecules show the
vibrational contributions. Although measurements at the same frequency are not
available for CHCl$_3$ and CDCl$_3$, Kaatz \textit{et al.}\cite{kaatz98} found that at 694.3
nm, CHCl$_3$ has $\beta_{\parallel}$ = 1.2 $\pm$ 2.6 au; at 1064 nm CDCl$_3$ 
has $\beta_{\parallel}$ = 1.0 $\pm$ 4.2 au. Given that the
frequency-dispersion of $\beta_{\parallel}$ between zero frequency and 1064 nm
is only about +15\% in our calculations (much smaller than the error bars), this
isotopic comparison shows that the vibrational contributions are less than the
error bars. Therefore vibrational contributions are not significant in comparing
the \textit{ab initio} results to the available experimental measurements. We
find additionally that the molecular structure has a significant influence on
the calculated value of $\beta_{\parallel}$, and so it is crucial to use an
accurate structure for theoretical calculations.

\begin{table}
\caption{
Summary of the best results obtained with the GTOs, numerical basis sets and
real-space grids, all using the PBE exchange-correlation functional. Real-space
grids (lr denotes linear response, and fd finite difference) have radius
22 $a_0$, spacing 0.25 $a_0$. All values are in atomic units.
}
\label{tab:Best_Res}
\begin{ruledtabular}
\begin{tabular}{lddddddddddd}
\tbhl{Basis Set} & \tbhc{$\mu_{z}$} & \tbhc{$\alpha_{yy}$} & \tbhc{$\alpha_{zz}$} & \tbhc{$\beta_{yyy}$} & \tbhc{$\beta_{yyz}$} & \tbhc{$\beta_{zzz}$} & \tbhc{$\bar{\alpha}$} & \tbhc{$\beta_{\parallel}$} & \tbhc{$\beta_{\mathrm{HRS}}^{\mathrm{VV}}$} \\
\hline
GTO d-aV5Zs    & 0.404 & 65.70 & 46.79 & 27.35 & -15.31 & 22.27 & 59.40 & -5.01 & 16.90 \\
NBS 5Z4Pe8     & 0.398 & 65.45 & 46.28 & 24.54 & -14.90 & 21.37 & 59.06 & -5.07 & 15.68 \\
RS lr          & 0.399 & 66.02 & 47.00 & 27.12 & -16.36 & 26.94 & 59.68 & -3.47 & 17.44 \\
\hline
RS fd          & \tbc{$''$} & 66.05 & 46.87 & 24.74 &  -15.17 &  23.89 &  59.66 &  -3.87 & 15.97 \\
RS 1064 nm     & \tbc{$''$} & 66.69 & 47.34 & 30.35 &  -18.95 &  31.56 &  60.24 &  -4.01 & 19.91 \\
\hline
\tbhl{Expt.} & \tbr{0.409$\pm$0.008}\cite{reinhart70} & \tbr{61$\pm$5}\cite{Stuart_Volkmann} & \tbr{45$\pm$3}\cite{Stuart_Volkmann} &  &  &  & \tbr{56$\pm$4}\cite{Stuart_Volkmann} & \tbr{1$\pm$4}\cite{kaatz98} & \\
\end{tabular}
\end{ruledtabular}
\end{table}

\begin{acknowledgments}

This work was supported in part by
U.S. Department of Energy Grant DE-FC36-08GO18024 (FV) and
DE-FG03-97ER45623 (JJR), and the National Science Foundation Grant
0120967 through the STC MDITR (YT and FV).  DS was supported by an
NSF IGERT fellowship.  This work was also supported by National Science
Foundation Grant No. DMR07-05941
and by the Director, Office of Science, Office of Basic Energy Sciences,
Materials Sciences and Engineering Division, U.S. Department of
Energy under Contract No. DE-AC02-05CH11231 (SGL). Computational resources have
been provided by DOE at Lawrence Berkeley National Laboratory's NERSC
facility and Lawrencium cluster.  We acknowledge funding by the
Spanish MEC (FIS2007-65702-C02-01),
``Grupos Consolidados UPV/EHU del Gobierno Vasco'' (IT-319-07) and the
European Community through  e-I3 ETSF project (Contract Number 211956).
We acknowledge support by the Barcelona Supercomputing Center, ``Red
Espa\~{n}ola de Supercomputaci\'{o}n,'' SGIker ARINA (UPV/EHU) and ACI-Promociona (ACI2009-1036).
\end{acknowledgments}

\bibliographystyle{apsrev}

\end{document}